\documentclass[aps,pre,reprint,superscriptaddress]{revtex4-2}
\usepackage{amsmath}
\usepackage{graphicx}          
\graphicspath{{figures/}}      

\usepackage{hyperref}
\hypersetup{
  colorlinks=true,    
  linkcolor=blue,     
  citecolor=blue,     
  urlcolor=blue,      
}
\begin{document}


\title{Bifurcation of the quasi-stationary velocity of strongly discrete transition waves driven by gravity}


\author{Zehuan Tang}
\affiliation{Department of Mechanics and Engineering Science, College of Civil Engineering and Mechanics, Lanzhou University, Lanzhou 730000, Gansu, China}

\author{Qing Xia}
\affiliation{School of Civil Engineering and Transportation, South China University of Technology, Guangzhou, 510641, Guangdong, China}

\author{Hui Chen}
\affiliation{School of Mechanical Engineering and Mechanics, Ningbo University, Ningbo 315211, Zhejiang, China}

\author{Songyang Fu}
\affiliation{Department of Mechanics and Engineering Science, College of Civil Engineering and Mechanics, Lanzhou University, Lanzhou 730000, Gansu, China}

\author{Yuanwen Gao}
\affiliation{Department of Mechanics and Engineering Science, College of Civil Engineering and Mechanics, Lanzhou University, Lanzhou 730000, Gansu, China}


\date{\today}

\begin{abstract}
Transition waves are common in multistable mechanical metamaterials, and the dynamics of weakly discrete transition waves under driving forces have been extensively discussed. However, as lattice effects become more pronounced, strongly discrete transition waves may exhibit dynamics that cannot be predicted by the continuum limit. Here, by tilting a bistable chain, we introduce a gravitational perturbation term into the dynamical equations, under which the transition waves are continuously accelerated. In the strongly discrete regime, we find that transition waves under gravitational driving possess quasi-stationary velocity plateaus (QSVPs), and the number of these plateaus first increases and then decreases as the tilt angle increases. We theoretically elucidate that the emergence of the velocity plateaus originates from the balance between gravitational driving and phonon radiation. In further analysis, the theoretical model reveals that the balance point undergoes a bifurcation at the radiation resonance, which leads to a change in the number of velocity plateaus. Our study extends the investigation of transition waves into the strongly discrete regime, and the emergence of multiple velocity plateaus opens up new possibilities for programmable solitary waves.
\end{abstract}


\maketitle
\section{INTRODUCTION}
Kink solitons, as fundamental soliton excitations in field theory, were realized in the Scott pendulum long ago~\cite{1}. In recent years, owing to the rapid development of mechanical metamaterials, kink solitons have been realized in an increasing number of solid structures. A intuitive feature of such solitons is that they can propagate in solid structures at non-zero velocity~\cite{2,3,4,5}, and also can exist stably in a static form~\cite{6,7,8,9}.

With the combination of soliton theory and topological band theory~\cite{10,11,12,13}, two types of kink solitons have emerged. One is the topologically protected kink soliton in the Kane-Lubensky chain~\cite{10,14}, often directly referred to as a topological soliton~\cite{15}. As a continuation of linear zero-energy modes, topological solitons naturally constitute models that are free of the static Peierls-Nabarro potential (PNp)~\cite{16}, and the continuum limit remains applicable within a certain range of strong discreteness~\cite{17}. The other type is the topologically trivial transition wave in bistable chains~\cite{18,19}. Transition waves cannot be reduced to any model of floppy mode, and the PNp already exists in the static case. Therefore, once one moves away from the weakly discrete regime, the dynamics of transition waves become difficult to predict.

In fact, existing studies on transition waves have largely been conducted within the weakly discrete regime, where many interesting conclusions have been drawn, such as the energy scaling law~\cite{20,21} and velocity evolution laws~\cite{22,23} under the adiabatic approximation. However, at the same time as strong discreteness violates the assumptions of the continuum approximation, it may also endow solitons with richer dynamics~\cite{24,25,26}. Here, by tilting a bistable chain, we systematically investigate the dynamics of strongly discrete transition waves under gravitational driving. Unlike the weakly discrete system where damping had to be introduced to achieve balance ~\cite{20,21,23}, we find that driving force can balance with the spontaneously radiated phonon modes of transition waves in strongly discrete system, and the number of such balanced modes first increases and then decreases as the perturbation magnitude grows. In theoretical model, this trend is explained by the variation in the number of intersections between the gravitational power curve and the radiation power curve. Our findings contribute to a more comprehensive understanding of the dynamic response of nonlinear mechanical metamaterials.

\section{BIASED BISTABLE CHAIN AND TRANSITION WAVES}

\begin{figure}[htbp]
\centering
\includegraphics[width=0.9\columnwidth]{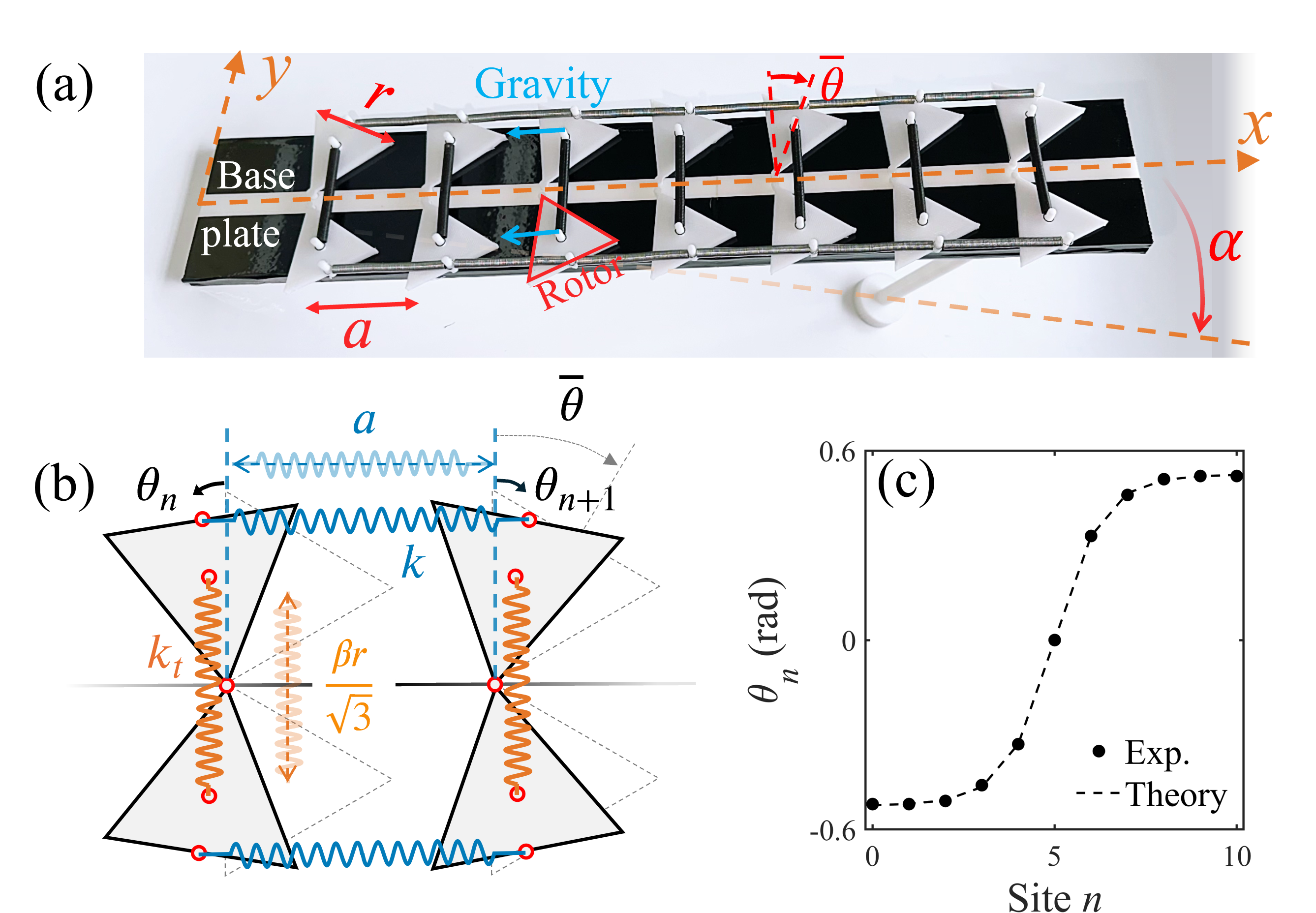}
\caption{\label{fig:Bistable chain}(a) A bistable chain with tilt angle $\alpha$. The angle $\overline{\theta}$ is the initial equilibrium angle of the bistable chain. (b) Schematic of the model. The dashed triangle represents the initial configuration, and the lengths of the light blue and light orange springs denote the natural lengths of the two springs. (c) Comparison of experimental and theoretical results for the static kink in a chain of 11 nodes. The experimental data were measured for $\alpha = 0$.}
\end{figure}

As shown in Fig.~\ref{fig:Bistable chain}(a), the two rows of rotors in the bistable chain are hinged on the base plate with a spacing $a$, where the equilateral triangular rotors have side length $r$, mass $m$, and moment of inertia $J$. The rotors are connected by linear springs. In the $x$-direction, the rotors are hinged at the side midpoints by springs of stiffness $k$, and this midpoint connection eliminates geometric asymmetry; in the $y$-direction, an additional spring of stiffness $k_{t}$ is attached between the mass centers of the rotors to create a double-well potential. Based on the modeling diagram in Fig.~\ref{fig:Bistable chain}(b), the potential energy of a unit of the bistable chain can be written as
\begin{equation}
V_{b}^{\,y} = \frac{1}{6} k_{t} r^{2} \left( 2\cos\theta_{n} - \beta \right)^{2}. \label{eq:Vby}
\end{equation}
The inter-unit coupling potential energy is
\begin{equation}
V_{b}^{\,x} = k \left( \sqrt{ \left( a + \Omega_{n}^{h} \right)^{2} + \left( \Omega_{n}^{v} \right)^{2} } - a \right)^{2}, \label{eq:Vbx}
\end{equation}
where $\Omega_{n}^{h} = \frac{\sqrt{3}r}{2}\left( \sin\theta_{n+1} - \sin\theta_{n} \right)$ and $\Omega_{n}^{v} = \frac{\sqrt{3}r}{2}\left( \cos\theta_{n+1} - \cos\theta_{n} \right)$ are the horizontal and vertical elongations of the blue spring, respectively. The geometric nonlinearity of structural deformation makes the expressions of $V_{b}^{\,y}$ and $V_{b}^{\,x}$ complicated, especially that of $V_{b}^{\,x}$, so we simplify them to obtain an approximate potential energy for the bistable chain. By performing a Taylor expansion of these expressions and neglecting higher-order terms, the potential energy of the bistable chain can be approximated as
\begin{equation}
V_{b} = \sum_{n} \left[ \frac{1}{6} k_{t} r^{2} \left( \theta_{n}^{2} - \overline{\theta}^{2} \right)^{2} + \frac{3}{4} k r^{2} \left( \theta_{n+1} - \theta_{n} \right)^{2} \right]. \label{eq:Vb}
\end{equation}
The static solution of the theoretical model is obtained from $\partial V_{b}/\partial\theta_{n} = 0$, and we have verified the correctness of our theoretical model through static experiments, as shown in Fig.~\ref{fig:Bistable chain}(c).

Next, we consider the dynamic model of the bistable chain in an ideal undamped environment. We tilt the base plate by an angle $\alpha$, so that gravity has a component in the $x$-direction. This gravitational component does not change the equilibrium angle of the structure, because we have chosen stiff springs satisfying $mg/k_{t} \ll 1$. Treating gravity as an external driving force of the system, the dimensionless Lagrangian of the bistable chain is written as
\begin{equation}
L = \ell \sum_{n} \left[ \frac{1}{2} \left( \frac{\partial \varphi_{n}}{\partial T} \right)^{2} - \left( \frac{\varphi_{n}^{2} - 1}{2} \right)^{2} - \left( \frac{\varphi_{n+1} - \varphi_{n}}{\sqrt{2} \ell} \right)^{2} \right], \label{eq:Lagrangian}
\end{equation}
where $\varphi_{n} = \theta_{n} / \overline{\theta}$ and $T = r\overline{\theta} \sqrt{k_{t}/(3J)}\, t$ are the normalized angle and time. $\ell = (2\overline{\theta}/3) \sqrt{k_{t}/k}$ is the spatial step size, which defines the discreteness of the system.

Considering the Euler-Lagrange equation with additional generalized forces,
\begin{equation}
\frac{\partial}{\partial T}\left( \frac{\partial\mathcal{L}_{n,n+1}}{\partial\dot{\varphi}_n} \right) - \frac{\partial\mathcal{L}_{n,n+1}}{\partial\varphi_n} - \frac{\partial\mathcal{L}_{n-1,n}}{\partial\varphi_n} = - \frac{\partial\widehat{F}_g}{\partial\varphi_n}, \label{eq:EulerLagrange}
\end{equation}
where $\widehat{F}_g = \frac{\sqrt{3}mg\sin\alpha}{k_t r \overline{\theta}^4} \sin(\overline{\theta}\varphi_n)$ is the dimensionless gravitational potential energy density. The dynamical equation of the bistable chain is written as
\begin{equation}
\frac{\partial^2\varphi_n}{\partial T^2} - \frac{(\Delta_s\varphi)_n}{\ell^2} - (1-\varphi_n^2)\varphi_n = -\epsilon \cos(\overline{\theta}\varphi_n), \label{eq:dyn}
\end{equation}
where $\Delta_s$ is the second-order difference operator defined by $(\Delta_s\varphi)_n = \varphi_{n+1} + \varphi_{n-1} - 2\varphi_n$; $\epsilon$ is the perturbation coefficient, $\epsilon \ll 1$, which can be tuned by the tilt angle $\alpha$.

\begin{figure}[htbp]
\centering
\includegraphics[width=0.95\columnwidth]{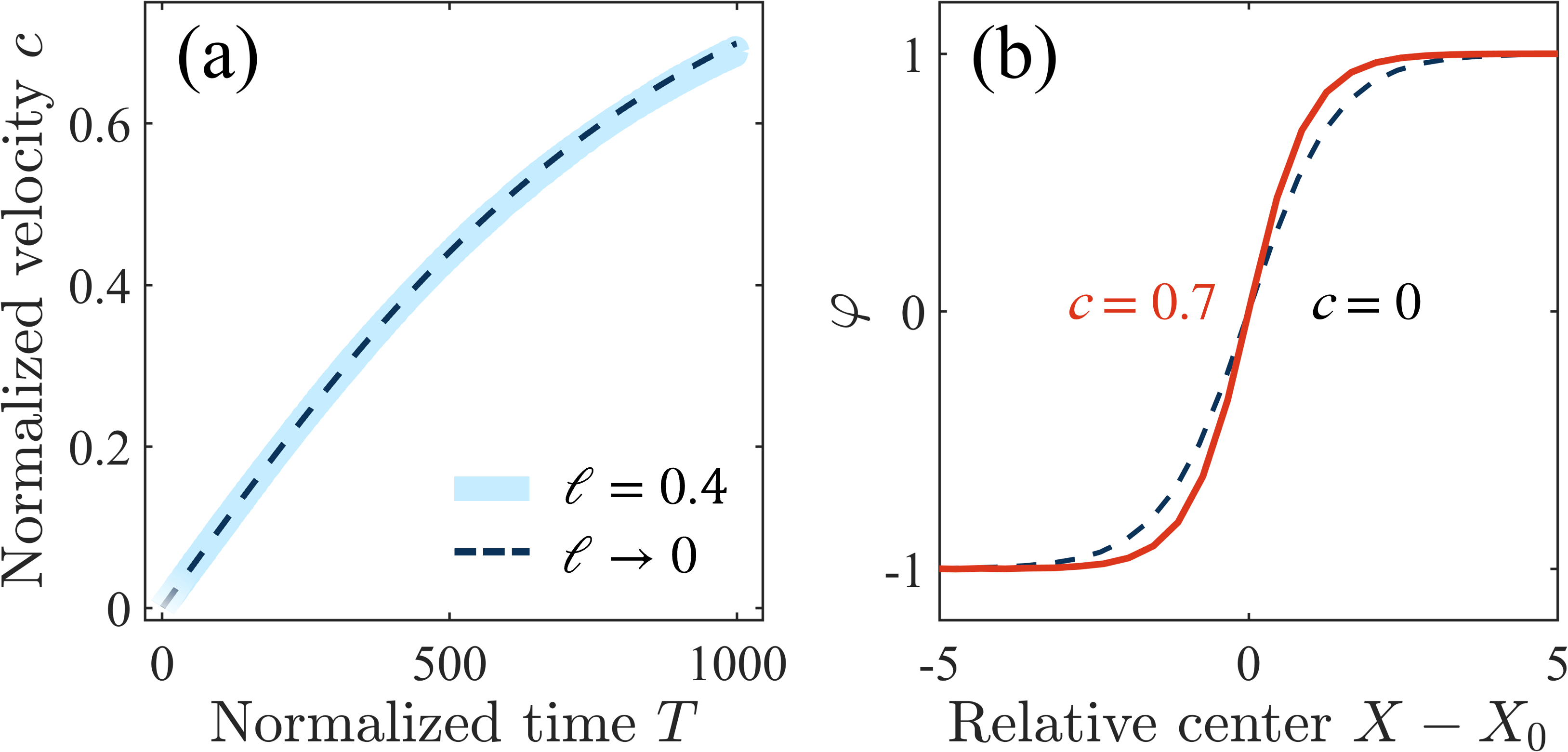}
\caption{\label{fig:weakly_discrete}(a) Comparison of acceleration curves for the weakly discrete case $\ell = 0.4$ and the continuum limit $\ell \rightarrow 0$. (b) The kink solution for $\ell = 0.4$ follows the framework of the continuum limit solution Eq.~\eqref{eq:kink}, with only Lorentz contraction occurring during acceleration.}
\end{figure}

\begin{figure*}[htbp]
\centering
\includegraphics[width=0.9\textwidth]{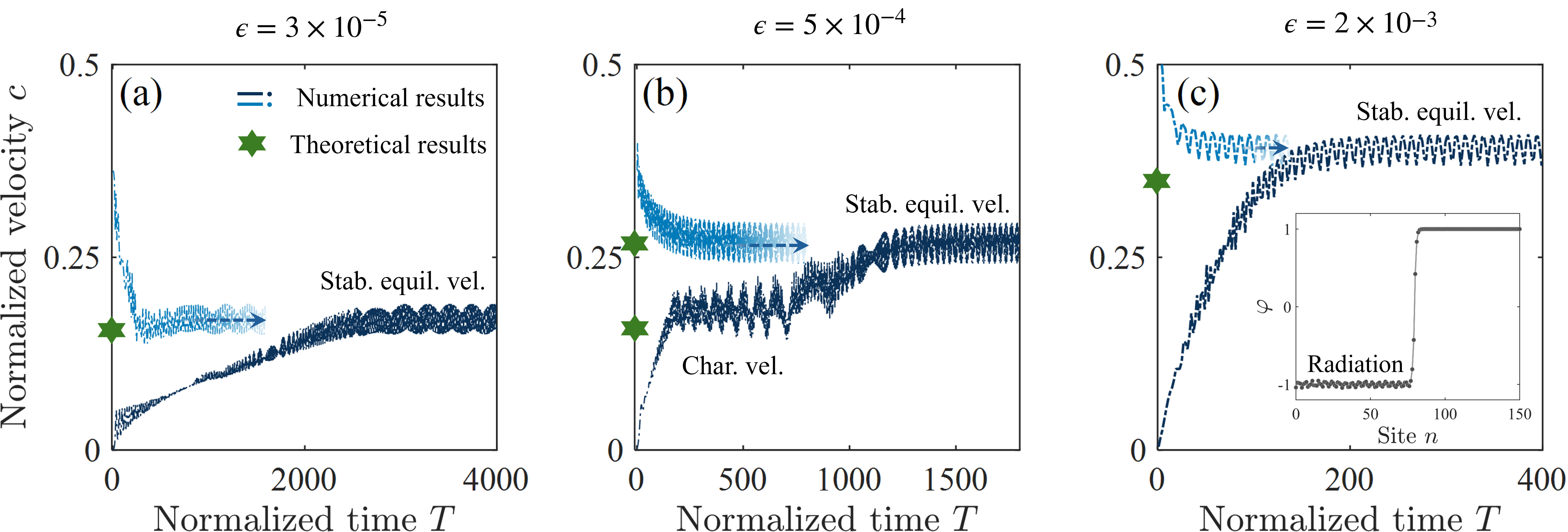}
\caption{\label{fig:strong_discrete}Evolution of the quasi-stationary velocity with the perturbation parameter. (a) Single velocity plateau curve for $\epsilon = 3 \times 10^{-5}$. The black velocity curve corresponds to excitation from zero initial velocity, while the blue curve corresponds to excitation from a larger initial velocity. (b) Double velocity plateau curve for $\epsilon = 5 \times 10^{-4}$. (c) Single velocity plateau curve for $\epsilon = 2 \times 10^{-3}$. The inset shows the phonon radiation at the wavefront. The theoretical results (cyan hexagons) in the three panels are obtained from Section III.C below.}
\end{figure*}

When $\ell$ is small, Eq.~\eqref{eq:dyn} can be approximated by the continuum limit as
\begin{equation}
\frac{\partial^2\varphi}{\partial T^2} - \frac{\partial^2\varphi}{\partial X^2} - (1-\varphi^2)\varphi = -\epsilon \cos(\overline{\theta}\varphi), \label{eq:continuum}
\end{equation}
which is a $\varphi^4$ equation with a perturbation term. Under the adiabatic approximation, the solution can be expressed as a kink that accelerates over time,
\begin{equation}
\varphi(X,T) = \tanh\frac{X - X_0(T)}{\sqrt{2\bigl(1 - (\dot{X}_0(T))^2\bigr)}}, \label{eq:kink}
\end{equation}
where $\dot{X}_0(T) = c(T)$ is the time-dependent velocity. In Fig.~\ref{fig:weakly_discrete}(a), taking the transition wave with $\ell = 0.4$ as an example, we numerically demonstrate that the acceleration dynamics in the weakly discrete regime can be approximated by the continuum limit. During the acceleration process, the transition wave only undergoes Lorentz contraction, with almost no phonon radiation being generated, as shown in Fig.~\ref{fig:weakly_discrete}(b).

Since the conditions for the continuum approximation are satisfied, the dynamics in the weakly discrete regime are simple and predictable. However, under strong discreteness, the breakdown of the continuum assumption makes the acceleration curve difficult to determine straightforwardly. In the following section, we systematically discuss the acceleration dynamics of strongly discrete transition waves.

\section{QUASI-STATIONARY VELOCITY PLATFORMS}
\subsection{Numerical results}

Without loss of generality, we first select the strongly discrete parameter $\ell = 1$ for discussion. In Fig.~\ref{fig:strong_discrete}, the black curve is the acceleration curve of the transition wave starting from zero initial velocity. This result shows that the transition wave expectedly deviates from the dynamics in the continuum limit, and it no longer possesses a smooth and monotonically accelerating curve as in Fig.~\ref{fig:weakly_discrete}(a). The generation of phonon radiation keeps the velocity of the transition wave oscillating continuously, while also forming QSVPs that lock the upper limit of acceleration process. As the perturbation coefficient increases, we observe that the velocity curve evolves from a single plateau to double plateaus and then back to a single plateau. Figs.~\ref{fig:strong_discrete}(a) –~\ref{fig:strong_discrete}(c) illustrate this evolution with examples of $\epsilon = 3 \times 10^{-5}$, $\epsilon = 5 \times 10^{-4}$ and $\epsilon = 2 \times 10^{-3}$.

In addition to the change in the number of velocity plateaus, we also observe that the two QSVPs in Fig.~\ref{fig:strong_discrete}(b) are fundamentally different. When the initial velocity is greater than the value of the second plateau, the blue velocity curve eventually converges to the second plateau instead of descending to the first plateau. From this observation, the second velocity plateau exhibits the characteristics of stable equilibrium: regardless of whether the initial velocity is above or below the stable equilibrium velocity, the velocity gradually converges to this stable equilibrium velocity. In comparison, the first velocity plateau behaves more like a characteristic velocity of the system. This is manifested by the fact that from Fig.~\ref{fig:strong_discrete}(b) to Fig.~\ref{fig:strong_discrete}(a), even though $\epsilon$ is reduced to $6\%$ of its original value, this velocity remains a quasi-stationary velocity of the system (though its stability changes to that of a stable equilibrium).

To better understand the above numerical results, in the following subsections we attempt to theoretically interpret them.

\subsection{Phonon radiation model}

\begin{figure}[htbp]
\centering
\includegraphics[width=0.8\columnwidth]{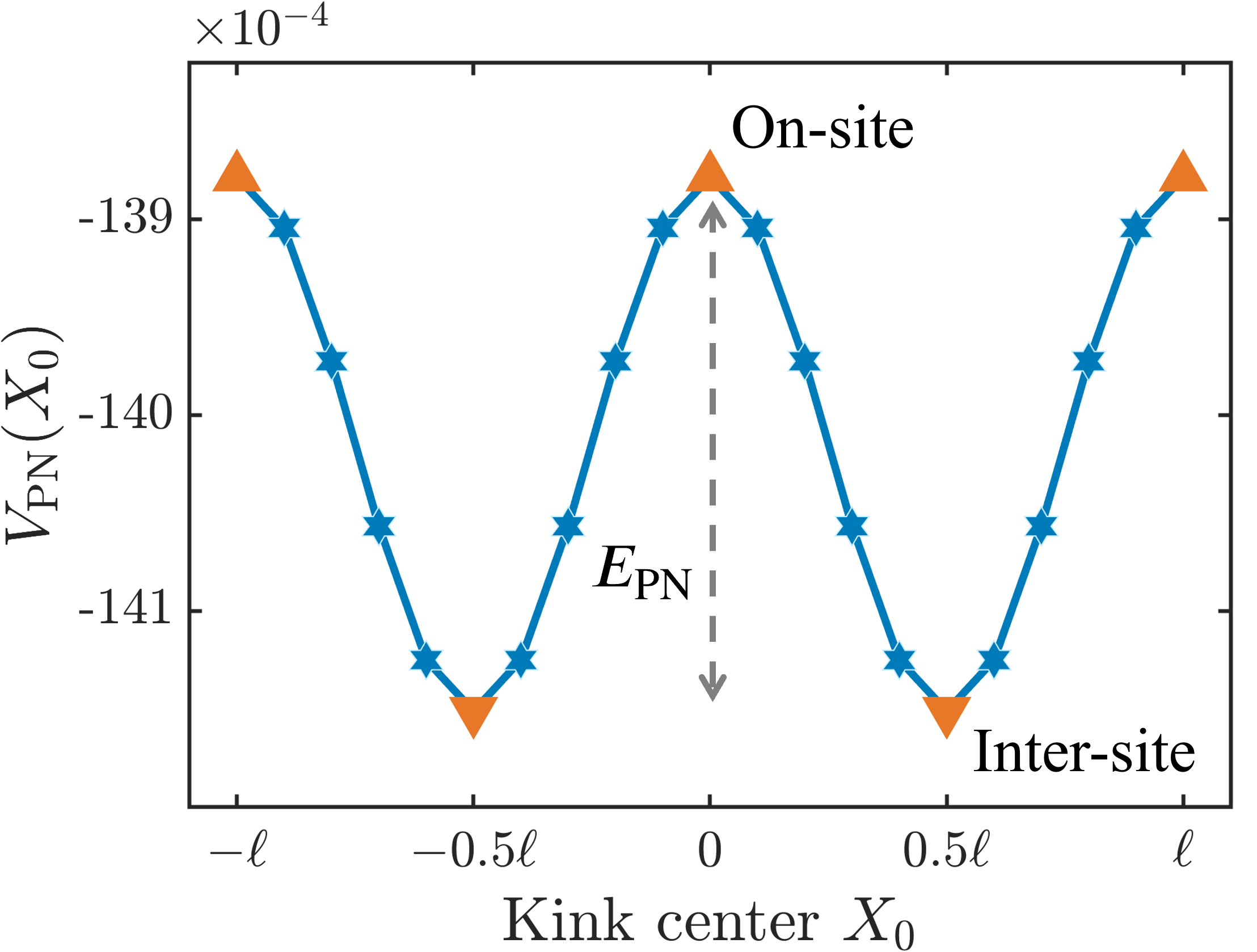} 
\caption{\label{fig:PNp}Periodically varying PNp. The upward-pointing triangles at the on-site positions correspond to the energy maxima of the system, and the downward-pointing triangles at the inter-site positions correspond to the energy minima. The difference between the maximum and minimum is the barrier height $E_{\text{PN}}$.}
\end{figure}

Before proceeding with theoretical modeling, we first revisit the concept of the PNp and introduce its modeling in our theory. The potential energy term in Eq.~\eqref{eq:Lagrangian} is rearranged as
\begin{align}
V = & \ell \sum_n \left\{ \frac{1}{2} \left( \frac{\varphi_{n+1} - \varphi_n}{\ell} \right)^2 + \frac{1}{4} \left( 1 - \frac{F(\varphi_n,\varphi_{n+1})}{3} \right)^2 \right\} \notag \\
& + \ell \sum_n \frac{1}{4} \left[ (1-\varphi_n^2)^2 - \left( 1 - \frac{F(\varphi_n,\varphi_{n+1})}{3} \right)^2 \right], \label{eq:V_rearranged}
\end{align}
where $F(\varphi_n,\varphi_{n+1}) = \varphi_{n+1}^2 + \varphi_n\varphi_{n+1} + \varphi_n^2$, and the sum of the products of the difference term and $F$, i.e., $\sum_n (\varphi_{n+1} - \varphi_n)F(\varphi_n,\varphi_{n+1})$, can be reduced to a boundary constant term. Therefore, the first part of Eq.~\eqref{eq:V_rearranged} constitutes the PNp‑free discrete form of the $\varphi^4$ model~\cite{27}, often also referred to as the topological discrete form~\cite{17}. Using the iterative solution $\{\varphi_n\}$ of the topological discrete form as the solution of the original discrete form, and denoting the boundary constant term by $C$, $V$ becomes
\begin{equation}
V = C + \ell \sum_n \frac{1}{4} \left[ (1-\varphi_n^2)^2 - \left( 1 - \frac{F(\varphi_n,\varphi_{n+1})}{3} \right)^2 \right]. \label{eq:V_const}
\end{equation}
The second part provides an expression for constructing the PNp, namely, $V_{\mathrm{PN}}$ can be written as
\begin{equation}
V_{\mathrm{PN}} = \ell \sum_n \frac{1}{4} \left[ (1-\varphi_n^2)^2 - \left( 1 - \frac{F(\varphi_n,\varphi_{n+1})}{3} \right)^2 \right]. \label{eq:V_PN}
\end{equation}
Fig.~\ref{fig:PNp} takes $\ell = 1$ as an example, showing the image of $V_{\mathrm{PN}}$. It can be seen that Eq.~\eqref{eq:V_PN} and the conventional construction $V_{\mathrm{PN}}(X_0) = \frac{E_\mathrm{PN}}{2} \left( \cos\frac{2\pi X_0}{\ell} - 1 \right)$ have the same period $\ell$. Moreover, it has the same physical meaning as the conventional construction: the on‑site case with $X_0/\ell = n$ corresponds to the energy maximum of the system, and the inter‑site case with $X_0/\ell = n + 1/2$ corresponds to the energy minimum. However, the explicit construction Eq.~\eqref{eq:V_PN} has a natural advantage, namely it can directly derive the expression for the force generated by the PNp, $f_n^{\mathrm{PN}} = - \frac{\partial(V_{\mathrm{PN}}/\ell)}{\partial\varphi_n}$. Its specific expression is
\begin{align}
f_n^{\mathrm{PN}} = & \frac{1}{18} \left[ (\varphi_n + \varphi_{n+1})^3 + (\varphi_n + \varphi_{n-1})^3 - 2(2\varphi_n)^3 \right] \notag \\
& - \frac{1}{6} (\varphi_{n+1} + \varphi_{n-1} - 2\varphi_n). \label{eq:fnPN_original}
\end{align}

\begin{figure}[htbp]
\centering
\includegraphics[width=0.95\columnwidth]{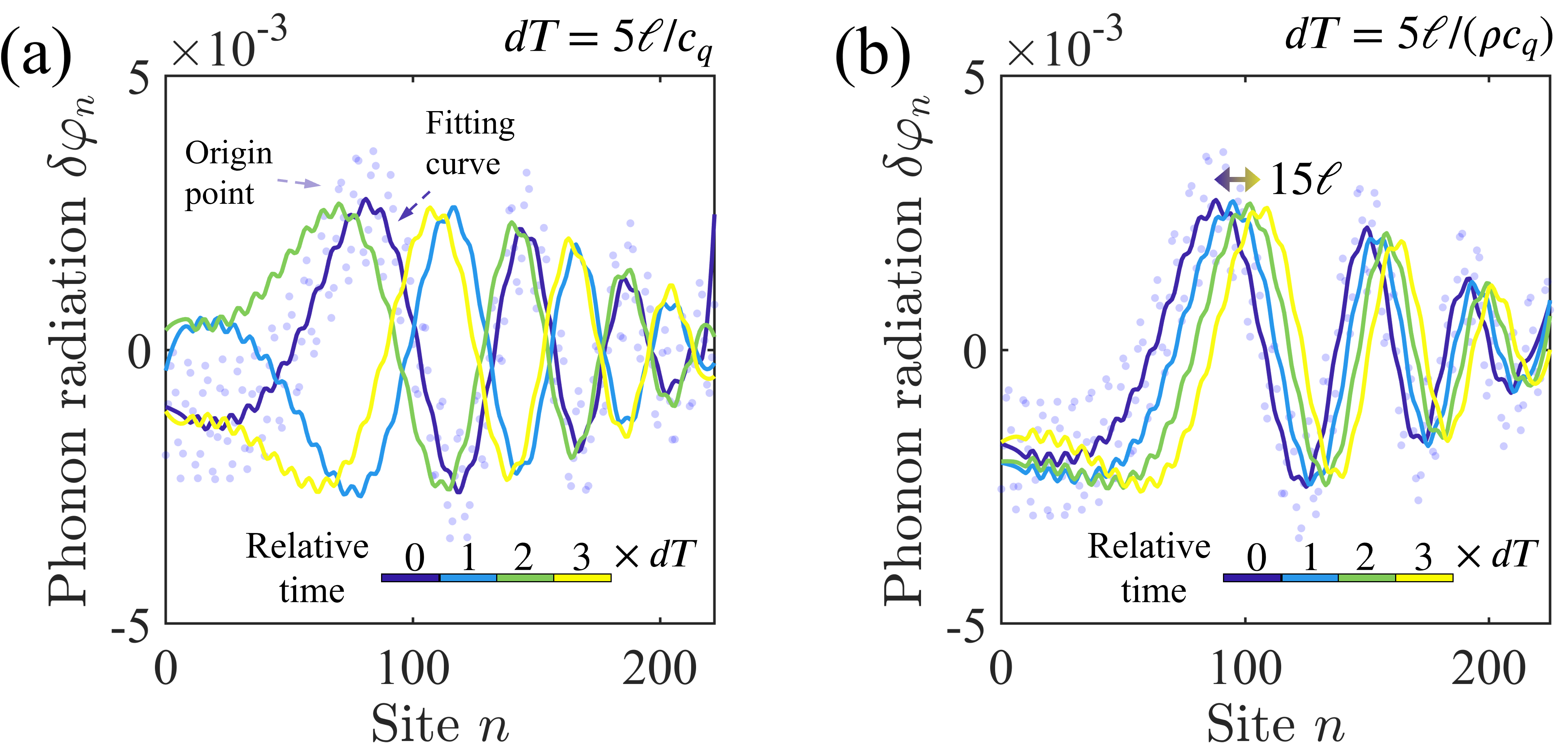} 
\caption{\label{fig:radiation}In the case of $\ell = 1$, the propagation diagram of phonon radiation at different time steps. (a) Propagation diagram of phonon radiation at time step of $\ell / c_q$. (b) Propagation diagram of phonon radiation at time step of $\ell / (\rho c_q)$, with $\rho$ set to 1.46.}
\end{figure}

It is a known fact that in PNp-free models, solitons hardly generate phonon radiation within a certain range of velocities. Therefore, the existence of PNp serves as the cause of the generation of phonon radiation, and we use the force $f_n^{\text{PN}}$ generated by PNp to simulate the excitation of phonon radiation. Namely
\begin{equation}
\frac{\partial^2\delta\varphi_n}{\partial T^2} - \frac{(\Delta_s\delta\varphi)_n}{\ell^2} + 2\delta\varphi_n = f_n^{\mathrm{PN}}(T), \label{eq:phonon_force}
\end{equation}
where $\delta\varphi_n$ is the radiation part of $\varphi_n$ in wavefront, $\varphi_n = -1 + \delta\varphi_n$. To further simplify Eq.~\eqref{eq:phonon_force}, its symmetry is discussed next. By observing the propagation of phonon radiation in the quasi-steady regime, we find that there exists a complicated coupling between the radiation and the soliton, namely the radiation does not satisfy the traveling-wave symmetry ${\delta\varphi}_{n+1}(T + \ell/c) = {\delta\varphi}_n(T)$ associated with the actual velocity of the soliton. Fig.~\ref{fig:radiation} demonstrates the breakdown of this symmetry: after measuring the quasi-stationary velocity $c_q$, when the time step is taken as $\ell/c_q$, $\delta\varphi_n(T)$ does not exhibit traveling-wave characteristics, as shown in Fig.~\ref{fig:radiation}(a); after introducing a coupling parameter $\rho$ to correct the symmetry, Fig.~\ref{fig:radiation}(b) shows that the phonon radiation regains traveling-wave characteristics with time step $\ell/(\rho c_q)$. This result indicates that the radiation of the $\varphi^4$ soliton should actually satisfy the symmetry
\begin{equation}
\delta\varphi_{n+1}\left( T + \frac{\ell}{\rho c} \right) = \delta\varphi_n(T), \label{eq:symmetry}
\end{equation}
and $\rho = 1.46$ is determined for $\ell = 1$. According to the symmetry of $\delta\varphi_n(T)$, applying the transformation $n \rightarrow n+1$ and $T \rightarrow T + \ell/(\rho c)$ simultaneously in Eq.~\eqref{eq:phonon_force} yields that $f_n^{\mathrm{PN}}(T)$ must also satisfy the symmetry
\begin{equation}
f_{n+1}^{\mathrm{PN}}\left( T + \frac{\ell}{\rho c} \right) = f_n^{\mathrm{PN}}(T). \label{eq:fn_symmetry}
\end{equation}
In the calculation framework of $f_n^{\mathrm{PN}}(T)$ provided by Eq.~\eqref{eq:fnPN_original}, the approximate expression of $f_n^{\mathrm{PN}}(T)$ can be obtained by substituting the continuum solution. Setting $\varphi_n(T) \approx \tanh\frac{n\ell - \rho cT}{W}$ and substituting it into Eq.~\eqref{eq:fnPN_original}, then $f_n^{\mathrm{PN}}(T)$ satisfying the symmetry requirement of Eq.~\eqref{eq:fn_symmetry} is obtained,
\begin{equation}
f_n^{\mathrm{PN}}(T) \approx \frac{\ell^2}{W^2} \tanh\frac{\zeta_n}{W} \left( 2\operatorname{sech}^4\frac{\zeta_n}{W} - \operatorname{sech}^2\frac{\zeta_n}{W} \right), \label{eq:fn_approx}
\end{equation}
where $\zeta_n = n\ell - \rho cT$ is the traveling-wave coordinate. At this point, the symmetries of $\delta\varphi_n(T)$ and $f_n^{\mathrm{PN}}(T)$ have been determined, and the establishment of the expression for $f_n^{\mathrm{PN}}(T)$ also makes Eq.~\eqref{eq:phonon_force} closed. Introducing a new time variable $\varGamma = T - \frac{n\ell}{\rho c}$, Eq.~\eqref{eq:phonon_force} is reduced to
\begin{equation}
\frac{\partial^2\delta\varphi_0(\varGamma)}{\partial\varGamma^2} - \frac{(\Delta_t\delta\varphi_0)(\varGamma)}{\ell^2} + 2\delta\varphi_0(\varGamma) = f_0^{\mathrm{PN}}(\varGamma), \label{eq:reduced}
\end{equation}
where $(\Delta_t\delta\varphi_0)(\varGamma) = \delta\varphi_0\!\left(\varGamma + \frac{\ell}{\rho c}\right) + \delta\varphi_0\!\left(\varGamma - \frac{\ell}{\rho c}\right) - 2\delta\varphi_0(\varGamma)$.

In Appendix B, we solve Eq.~\eqref{eq:reduced} and obtain
\begin{equation}
\delta\varphi_0(\varGamma) = \sum_{k>0} A_k \cos(\omega_k \varGamma), \label{eq:delta_solution}
\end{equation}
The Hamiltonian of the radiation wave is written as
\begin{equation}
H = \ell \sum_n \left[ \frac{1}{2} \left( \frac{\partial\delta\varphi_n}{\partial\varGamma} \right)^2 + \frac{1}{2} \left( \frac{\delta\varphi_n - \delta\varphi_{n+1}}{\ell} \right)^2 + \delta\varphi_n^2 \right]. \label{eq:Hamiltonian}
\end{equation}
Considering the energy carried by the radiation wave at the $n$-th node as the energy lost when the transition wave propagates from the $n$-th node to the $(n+1)$-th node, the lost energy can be written as $\Delta E = \mathcal{H} \ell$. After taking the time average of $\mathcal{H}$, the expression for the lost energy becomes $\Delta E = \frac{\ell}{2} \sum_{k>0} A_k^2 \omega_k^2$. Converted to the time scale, the time for the transition wave to propagate from the $n$-th node to the $(n+1)$-th node is $\Delta T = \ell / c$, and the power of energy loss caused by radiation is
\begin{equation}
P_r = \frac{\Delta E}{\Delta T} = \frac{c}{2} \sum_{k>0} A_k^2 \omega_k^2. \label{eq:Pr}
\end{equation}

In the modeling of this subsection, we obtain the radiation power curve $P_r$, which is plotted as the dark blue curve in Fig.~\ref{fig:power_balance}. But the radiation power curve alone cannot fully explain the results in Fig.~\ref{fig:strong_discrete}, because the system is also subject to gravity. Therefore, in the following subsection, we model the gravitational power, thereby comprehensively considering the joint action of radiation and gravity.

\subsection{Power balance}

We calculate the power of the gravitational perturbation term in the continuum limit. The Hamiltonian density corresponding to Eq.~\eqref{eq:continuum} is
\begin{equation}
\mathcal{H} = \frac{1}{2} \left( \frac{\partial\varphi}{\partial T} \right)^2 + \frac{1}{2} \left( \frac{\partial\varphi}{\partial X} \right)^2 + \frac{1}{4} (\varphi^2 - 1)^2, \label{eq:hamiltonian_density}
\end{equation}
The energy of the soliton is $E = \int_{-\infty}^{+\infty} \mathcal{H} \, dX$, and its rate of change can be expressed via the perturbation term as
\begin{equation}
\frac{dE}{dT} = \int_{-\infty}^{+\infty} \frac{\partial\mathcal{H}}{\partial T} \, dX = \int_{-\infty}^{+\infty} \left[ -\epsilon \cos(\overline{\theta}\varphi) \right] \frac{\partial\varphi}{\partial T} \, dX. \label{eq:energy_rate}
\end{equation}
Since $\overline{\theta}$ is usually designed to be a small angle (for example, $\overline{\theta} = 0.5\ \text{rad}$ in our structure), $\cos(\overline{\theta}\varphi)$ is expanded to second order as $\cos(\overline{\theta}\varphi) = 1 - \frac{1}{2} (\overline{\theta}\varphi)^2$ to estimate the power,
\begin{equation}
P_g = \frac{6 - \overline{\theta}^2}{3} \epsilon c. \label{eq:Pg}
\end{equation}
The total power of the system is $\Delta P = P_g - P_r$, and the intersection of $P_g$ and $P_r$ gives the equilibrium velocity of the system.

\begin{figure*}[t]
\centering
\includegraphics[width=0.7\textwidth]{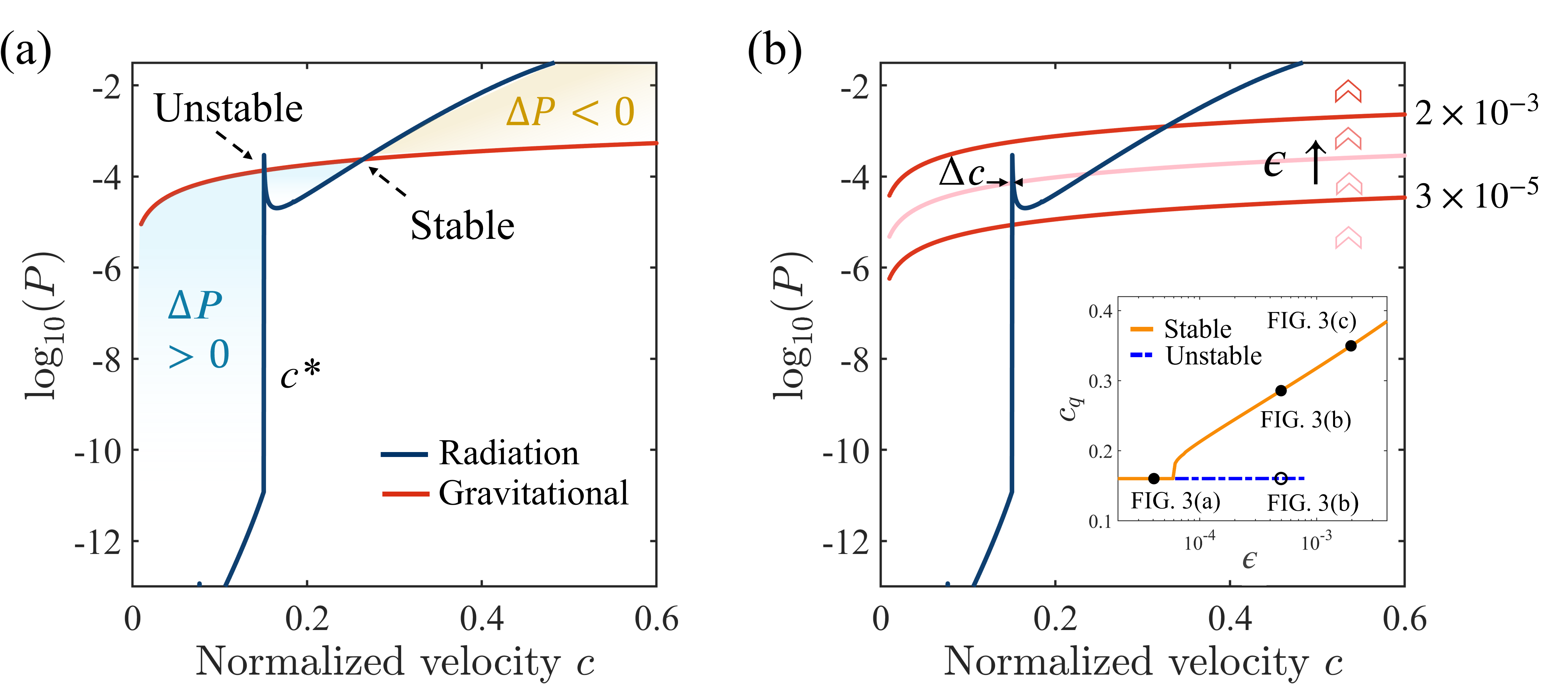}
\caption{\label{fig:power_balance}(a) Diagram illustrating the formation mechanism of the double velocity plateau. The first intersection of $P_g$ and $P_r$ is unstable. Because the actual velocity oscillates, this point will eventually repel the velocity to the right. The second intersection is stable, and it always attracts the velocity to its vicinity. (b) Evolution of the intersections with the parameter $\epsilon$. The inset is the bifurcation diagram of the equilibrium velocity versus the parameter $\epsilon$. Since the amplitude of the velocity oscillations is usually much larger than the peak width, the two intersections in the peak region are equivalent to one unstable equilibrium point.}
\end{figure*}

Fig.~\ref{fig:power_balance}(a) shows the variation of $\left. P_g \right|_{\epsilon = 5 \times 10^{-4}}$ and $P_r$ with the velocity $c$, where $\left. P_g \right|_{\epsilon = 5 \times 10^{-4}}$ is smooth, while $P_r$ exhibits a jump at the resonance velocity (denoted as $c^{*}$ in the figure). The two curves have two intersection velocities, $c^{*} = 0.15$ and $0.27$. When they are plotted as hexagons in Fig.~\ref{fig:strong_discrete}(b), the intersection velocities and the plateau velocities show fairly good agreement, indicating that the velocity plateau in Fig.~\ref{fig:strong_discrete}(b) is generated by the equilibrium point of the system. The consistent results also confirm our initial estimate. The first quasi-stationary velocity indeed corresponds to a characteristic velocity of the system, namely the resonance velocity $c^{*} = 0.15$. But the power curves reveal a clearer physical mechanism, namely that the first QSVP is formed by the intersection of the gravitational power curve and the radiation power curve at the resonant peak. On both sides of this intersection the total power is greater than zero. This intersection is an unstable equilibrium point, and under a sustained perturbation the velocity will be repelled to the right and eventually away from this point. This explains the phenomenon that the transition wave in Fig.~\ref{fig:strong_discrete}(b) eventually accelerates away from the first velocity plateau. The intersection of the gravitational power curve with the smooth segment of the radiation power curve is precisely the stable equilibrium point, i.e., the intersection at $0.27$, and the velocity will eventually always be stabilized at this point. This is the reason that, in Fig.~\ref{fig:strong_discrete}(b), even if the initial velocity exceeds the second quasi-stationary velocity, the velocity still returns to that second quasi-stationary velocity.

In Fig.~\ref{fig:power_balance}(b), two dark pink curves are used to show the geometric relationship between $\left. P_g \right|_{\epsilon = 3 \times 10^{-5}}$ and $\left. P_g \right|_{\epsilon = 2 \times 10^{-3}}$ with $P_r$. The intersection patterns of $\left. P_g \right|_{\epsilon = 3 \times 10^{-5}}$ and $\left. P_g \right|_{\epsilon = 2 \times 10^{-3}}$ with $P_r$ illustrate that, although both Fig.~\ref{fig:strong_discrete}(a) and Fig.~\ref{fig:strong_discrete}(c) show a single velocity plateau, they arise from equilibrium points formed under different scenarios. The quasi-stationary velocity in Fig.~\ref{fig:strong_discrete}(a) is formed by the intersection of $P_g$ with the jump segment of $P_r$. This quasi-stationary velocity does not change with $\epsilon$, so it remains nearly the same as the first quasi-stationary velocity in Fig.~\ref{fig:strong_discrete}(b). In contrast, the quasi-stationary velocity in Fig.~\ref{fig:strong_discrete}(c) is formed by the intersection of $P_g$ with the smooth segment of $P_r$. This quasi-stationary velocity necessarily increases as $\epsilon$ increases, which corresponds to the fact that the quasi-stationary velocity in Fig.~\ref{fig:strong_discrete}(c) is significantly larger than the second quasi-stationary velocity in Fig.~\ref{fig:strong_discrete}(b). In addition, the arrows in Fig.~\ref{fig:power_balance}(b) indicate that as the parameter $\epsilon$ increases, $P_g$ shifts upward. By scanning the parameters, we plot the variation of the equilibrium velocity with the parameter $\epsilon$, as shown in the inset of Fig.~\ref{fig:power_balance}(b). As can be seen from the inset, in Fig.~\ref{fig:strong_discrete} the non-monotonic evolution of the plateau number with the parameter $\epsilon$ can actually be attributed to a bifurcation behavior. After the bifurcation point, the $c = 0.15$ branch changes from a stable equilibrium to an unstable equilibrium, which corresponds to the transition of the stability of the quasi-stationary velocity from Fig.~\ref{fig:strong_discrete}(a) to Fig.~\ref{fig:strong_discrete}(b).

\section{THE MORE EXTENSIVE CASE OF $\ell$}

\begin{figure}[htbp]
\centering
\includegraphics[width=0.95\columnwidth]{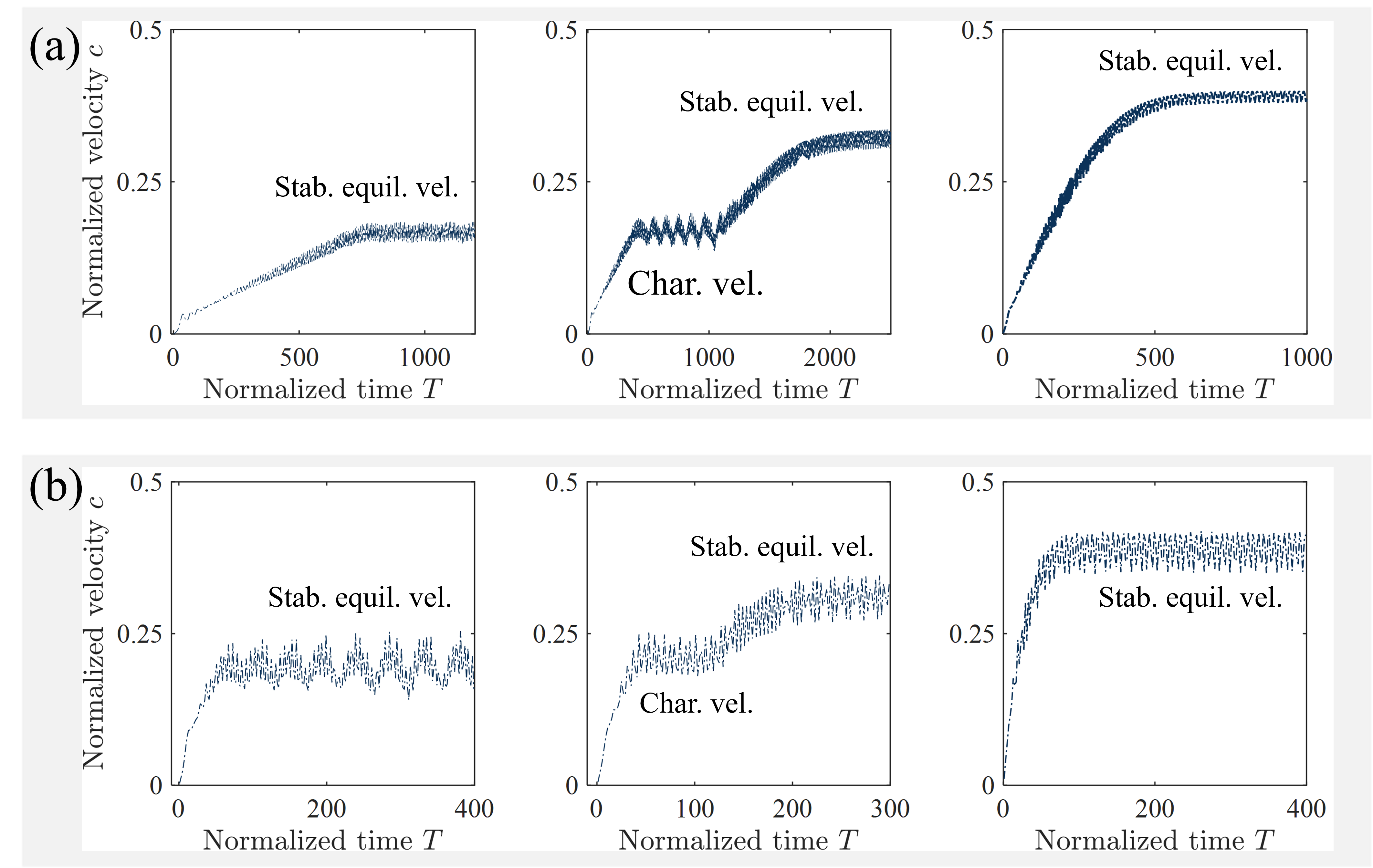}
\caption{\label{fig:extensive}(a) Evolution curves of the velocity for different $\epsilon$ in the discrete system with $\ell = 0.9$. (b) Evolution curves of the velocity for different $\epsilon$ in the discrete system with $\ell = 1.1$.}
\end{figure}

In this section, we aim to show that the phenomenon of the evolution of the number of velocity plateaus is not limited to the case $\ell = 1$ discussed above, but exists widely in strongly discrete systems. In Figs.~\ref{fig:extensive}(a) and Fig.~\ref{fig:extensive}(b), we show that the same evolution phenomena occur for $\ell = 0.9$ and $\ell = 1.1$, and they can also be interpreted within the above theoretical framework.

\section{CONCLUSIONS}

In summary, we numerically and theoretically studied the dynamics of strongly discrete transition waves under gravitational driving. Our results show that: under a small perturbation, the transition wave possesses only one stable quasi-stationary velocity at the radiation resonance; under a moderate perturbation, the stability of the quasi-stationary velocity at the resonance transitions to unstable, and a new stable quasi-stationary velocity appears at a higher velocity; under a large perturbation, the quasi-stationary velocity at the resonance disappears, and the system has only a stable quasi-stationary velocity far away from the resonance velocity. The theoretical analysis interprets this evolution as the change of equilibrium points caused by the shift of the gravitational power curve, and identifies the radiation resonance as the key to the bifurcation of the quasi-stationary velocity. Under this resonance condition, we anticipate that such bifurcation phenomena may also occur in other strongly discrete systems, such as rotating square chains~\cite{28,29} and bistable rotor chains~\cite{24}.

\section*{Acknowledgments}
This work was supported by the National Natural Science Foundation of China (Nos. 12272154 and 12532010).

\section*{DATA AVAILABILITY}
Data that support the findings of this article are not publicly available. The data are available from the authors upon reasonable request.

\appendix
\section{THE APPROXIMATION OF $f_n^{\mathrm{PN}}(T)$}
\renewcommand{\theequation}{\Alph{section}\arabic{equation}}
\setcounter{equation}{0}

Decompose Eq.~\eqref{eq:fnPN_original} into
\begin{align}
f_n^{\mathrm{PN}} = & \frac{(\varphi_{n+1} - \varphi_n)}{18}
\bigl[(\varphi_n+\varphi_{n+1})^2+2\varphi_n(\varphi_n+\varphi_{n+1})\bigr] \notag \\
& + \frac{(\varphi_{n-1} - \varphi_n)}{18}
\bigl[(\varphi_n+\varphi_{n-1})^2+2\varphi_n(\varphi_n+\varphi_{n-1})\bigr] \notag \\
& + \left(\frac{2}{9}\varphi_n^2-\frac16\right)
(\varphi_{n+1}+\varphi_{n-1}-2\varphi_n). \label{eq:A1}
\end{align}
This expression is complicated. Next, we organize it into a relatively concise approximate expression.

Since the continuum solution is analytic, we use $\varphi_n(T) = \varphi(X,T)\big|_{X = n\ell}$ to approximate $\varphi_n$ in Eq.~\eqref{eq:A1}. However, there is an error in this approximation caused by the deviation between the discrete solution and the continuum solution, which can be quantified as
\begin{equation}
R = \sum_n \bigl( \varphi_n - \varphi(n\ell,T) \bigr)^2. \label{eq:A2}
\end{equation}
To minimize the error of this approximation, a discrete contraction factor $\eta$ is introduced into the continuum solution to better approach the discrete solution, namely
\begin{equation}
\varphi(X,T) = \tanh\frac{X - \rho cT}{\eta\sqrt{2\bigl(1-(\rho c)^2\bigr)}}. \label{eq:A3}
\end{equation}
$\eta$ is determined when $R$ reaches its minimum. For the discreteness parameter $\ell = 1$ that is the focus of this paper, $\eta$ is determined to be $0.95$ in the static case. In dynamics, we still use the $\eta$ determined in the static case to reduce the error, namely, $\eta$ in Eq.~\eqref{eq:A3} is fixed at $0.95$.

Above, $\varphi_n$ is interpolated as a continuous function Eq.~\eqref{eq:A3}, where $\varphi_{n\pm 1}$ can be Taylor expanded as
\begin{equation}
\varphi_{n\pm 1} = \varphi_n + \sum_{m=1}^{\infty} \frac{(\pm \ell)^m}{m!} \frac{\partial^m \varphi}{\partial X^m}\bigg|_{X = n\ell}. \label{eq:A4}
\end{equation}
Substituting Eq.~\eqref{eq:A4} into Eq.~\eqref{eq:A1}, $f_n^{\mathrm{PN}}$ is rearranged as a series in $\ell$,
\begin{equation}
f_n^{\mathrm{PN}} = \sum_{m=1}^{\infty} S^{(m)} \ell^{m}, \label{eq:A5}
\end{equation}
where $S^{(m)}$ are the coefficients determined by the partial derivatives of $\varphi(X,T)$. In Eq.~\eqref{eq:A5}, the larger the step size $\ell$, the higher the order required when using the continuum solution to approximate $f_n^{\mathrm{PN}}$. For the theoretical solution of the parameter $\ell = 1$, we perform the solution to third-order accuracy,
\begin{equation}
f_n^{\mathrm{PN}} = S^{(1)} \ell + S^{(2)} \ell^{2} + S^{(3)} \ell^{3} + \mathcal{O}(\ell^{4}). \label{eq:A6}
\end{equation}
Based on the result of substituting Eq.~\eqref{eq:A4} into Eq.~\eqref{eq:A1}, the coefficients are determined as
\begin{align}
S^{(1)} &= S^{(3)} = 0, \notag \\
S^{(2)} &= \frac{2}{3} \varphi \left( \frac{\partial\varphi}{\partial X} \bigg|_{X = n\ell} \right)^2 + \left( \frac{2}{3}\varphi^2 - \frac{1}{6} \right) \frac{\partial^{2}\varphi}{\partial X^{2}} \bigg|_{X = n\ell}. \label{eq:A7}
\end{align}
In Eq.~\eqref{eq:A6}, substituting Eq.~\eqref{eq:A3} into $S^{(2)}$ and neglecting the truncation error $\mathcal{O}(\ell^{4})$ yields
\begin{equation}
f_n^{\mathrm{PN}}(T) \approx \frac{\ell^2}{W^2} \tanh\frac{\zeta_n}{W} \left( 2\operatorname{sech}^4\frac{\zeta_n}{W} - \operatorname{sech}^2\frac{\zeta_n}{W} \right), \label{eq:A8}
\end{equation}
where $\zeta_n = n\ell - \rho cT$ and $W = \eta\sqrt{2\bigl(1-(\rho c)^2\bigr)}$.

\section{SOLUTION OF THE RADIATION EQUATION}
\renewcommand{\theequation}{\Alph{section}\arabic{equation}}
\setcounter{equation}{0}

Applying a Fourier transform to Eq.~\eqref{eq:reduced} yields
\begin{equation}
\widetilde{\delta\varphi}_0(\omega) = \frac{\widetilde{f}_0^{\mathrm{PN}}(\omega)}{-\omega^2 - \frac{2}{\ell^2}\cos\frac{\omega\ell}{\rho c} + \left(2 + \frac{2}{\ell^2}\right)}. \label{eq:B1}
\end{equation}
$\widetilde{f}_0^{\mathrm{PN}}(\omega)$ is the Fourier transform of $f_0^{\mathrm{PN}}(T)$,
\begin{equation}
\widetilde{f}_0^{\mathrm{PN}}(\omega) = i\frac{\pi\omega^2\left( \frac{1}{12}\frac{\omega^2}{\lambda^2} - \frac{1}{6} \right)}{\lambda^3 W^2 \sinh\left( \frac{\pi\omega}{2\lambda} \right)} \ell^2, \label{eq:B2}
\end{equation}
where $\lambda = \rho c / W$. Its solution uses the following results:
\begin{equation}
\mathcal{F}\left\{ \tanh(\lambda\varGamma) \operatorname{sech}^2(\lambda\varGamma) \right\} = -i\frac{\pi\omega^2}{2\lambda^3} \frac{1}{\sinh\left( \frac{\pi\omega}{2\lambda} \right)}, \label{eq:B3}
\end{equation}
\begin{equation}
\mathcal{F}\left\{ \tanh(\lambda\varGamma) \operatorname{sech}^4(\lambda\varGamma) \right\} = -i\frac{\pi\omega^2}{24\lambda^3} \frac{\left(4 + \frac{\omega^2}{\lambda^2}\right)}{\sinh\left( \frac{\pi\omega}{2\lambda} \right)}. \label{eq:B4}
\end{equation}
The denominator of Eq.~\eqref{eq:B1} gives the dispersion relation of phonon radiation,
\begin{equation}
D(\omega,\rho c) = -\omega^2 - \frac{2}{\ell^2}\cos\frac{\omega\ell}{\rho c} + \left(2 + \frac{2}{\ell^2}\right) = 0.
\label{eq:B5}
\end{equation}
Let the roots of the dispersion equation be denoted as $\omega_k$. The number of $\omega_k$ determines the number of allowed phonon radiation modes in the system~\cite{30}.

Applying the inverse Fourier transform to $\widetilde{\delta\varphi}_0(\omega)$ recovers the time-domain solution,
\begin{equation}
\mathcal{F}^{-1}\left\{ \widetilde{\delta\varphi}_0(\omega) \right\} = i \sum_k \operatorname{Res}\left( \frac{\widetilde{f}_0^{\mathrm{PN}}(\omega)}{D(\omega)} e^{i\omega\varGamma},\ \omega_k \right). \label{eq:B6}
\end{equation}
Computing the residues gives
\begin{equation}
\delta\varphi_0(\varGamma) = \sum_{k>0} A_k \cos(\omega_k \varGamma), \label{eq:B7}
\end{equation}
where $A_k = \dfrac{2i\widetilde{f}_0^{\mathrm{PN}}(\omega_k)}{\partial_\omega D(\omega_k)}$ is the amplitude of the radiation wave.
\bibliography{myres}

\end{document}